\documentclass[iop,revtex4]{emulateapj}

\shortauthors{LAPI ET AL.}
\shorttitle{GALAXY EVOLUTION AT HIGH Z}

\begin{document}

\title{Galaxy Evolution at High Redshift:\\ Obscured Star Formation, GRB Rates, Cosmic Reionization, and Missing Satellites}
\author{A. Lapi\altaffilmark{1,2,3}, C. Mancuso\altaffilmark{1,2,3}, A. Celotti\altaffilmark{1,2,3}, L. Danese\altaffilmark{1,2,3}}
\altaffiltext{1}{SISSA, Via Bonomea 265, 34136 Trieste, Italy}
\altaffiltext{2}{INAF-Osservatorio Astronomico di Trieste, via Tiepolo 11, 34131 Trieste, Italy} \altaffiltext{3}{INFN-Sezione di
Trieste, via Valerio 2, 34127 Trieste, Italy}

\begin{abstract}
We provide an holistic view of galaxy evolution at high redshift $z\ga 4$, that incorporates the constraints from various astrophysical/cosmological probes, including the estimate of the cosmic SFR density from UV/IR surveys and long GRB rates, the cosmic reionization history after the latest \textsl{Planck} measurements, and the missing satellites issue. We achieve this goal in a model-independent way by exploiting the SFR functions derived by Mancuso et al. (2016) on the basis of an educated extrapolation of the latest UV/far-IR data from \textsl{HST}/\textsl{Herschel}, and already tested against a number of independent observables.  Our SFR functions integrated down to an UV magnitude limit $M_{\rm UV}\la -13$ (or SFR limit around $10^{-2}\, M_\odot$ yr$^{-1}$) produces a cosmic SFR density in excellent agreement with recent determinations from IR surveys and, taking into account a metallicity ceiling $Z\la Z_\odot/2$, with the estimates from long GRB rates. They also yield a cosmic reionization history consistent with that implied by the recent measurements of the \textsl{Planck} mission on the electron scattering optical depth $\tau_{\rm es}\approx 0.058$; remarkably, this result is obtained under a conceivable assumption regarding the average value $f_{\rm esc}\approx 0.1$ of the escape fraction for ionizing photons. We demonstrate via the abundance matching technique that the above constraints concurrently imply galaxy formation to become inefficient within dark matter halos of mass below a few $10^8\, M_\odot$; pleasingly, such a limit is also required not to run into the missing satellite issue. Finally, we predict a downturn of the galaxy luminosity function faintward of $M_{\rm UV}\la -12$, and stress that its detailed shape, as plausibly probed in the next future by the \textsl{JWST}, will be extremely informative on the astrophysics of galaxy formation in small halos, or even on the microscopic nature of the dark matter.
\end{abstract}


\setcounter{footnote}{0}

\section{Introduction}\label{sec|intro}

Recent observations of the high-redshift Universe have substantially improved our knowledge of the early stages in galaxy formation and evolution.

On the astrophysical side, UV observations from \textsl{HST} (e.g., Bouwens et al. 2015, 2016a,b; Finkelstein et al. 2015) and far-IR observations from \textsl{Herschel} surveys (e.g., Lapi et al. 2011; Gruppioni et al. 2013, 2015) have allowed to estimate the cosmic star formation rate (SFR) density (see Schiminovich et al. 2005; Hopkins \& Beacom 2006; Madau \& Dickinson 2014; Rowan-Robinson et al. 2016) and even to infer the shape of the galaxy SFR distributions (see Mancuso et al. 2016), including the essential contribution from strongly starforming dust-obscured objects, out to redshift $z\la 10$. Independent measurements have been also provided by estimates of the long gamma-ray bursts (GRB) rates from \textsl{Swift} observations (see Kistler et al. 2009, 2013; Chary et al. 2016 and references therein), that can effectively probe larger volumes than UV and far-IR surveys unbiasedly with respect to dust extinction, though being still affected by appreciable uncertainties.

On the cosmological side, the history of cosmic reionization has been recently probed to an unprecedented accuracy by the \textsl{Planck} Collaboration XLVII (2016) in terms of the optical depth for electron scattering $\tau_{\rm es}\approx 0.058$. Such data, besides assumptions concerning the escape fraction of ionizing photons from the early galaxies, provide independent constraints on the shape of the SFR function at high redshift.

A substantial, if not major, contribution to the cosmic SFR density at high redshift $z\ga 6$ and to the ionizing background responsible for reionization, comes from faint galaxies residing in small dark matter halos.  Numerical simulations indicate that an appreciable number of such small halos would survive down to the present time as bound satellites of Milky Way-sized galaxies, which are instead not observed in the local Universe. This constitutes the missing satellite problem, one of the most serious issue faced by the standard $\Lambda$CDM model (see Boylan-Kolchin et al. 2014; Wetzel et al. 2016). It can be solved by invoking astrophysical processes that must severely limit or even suppress galaxy formation in halos with masses of a few $10^8\, M_\odot$, or alternatively by abandoning the paradigm on the 'cold' microscopic nature of the dark matter (e.g., Lapi \& Danese 2015). Whatever the solution is, such an argument establishes an intriguing connection between high-redshift constraints and local observables.

Here we aim at providing an holistic view of galaxy evolution at high redshift $z\ga 4$ by exploiting jointly the above astrophysical and cosmological probes; remarkably, we will demonstrate that a coherent picture emerges, and we will provide specific predictions to test it with further observations in the next future. The plan of the paper is the following: in \S~2 we focus on the cosmic SFR history, as inferred from UV/IR surveys and as estimated from GRB rates; in \S~3 we consider the cosmic reionization history as probed by the recent data from \textsl{Planck} mission; in \S~4 we establish a connection between the high-redshift observables and the missing satellite issue in the local Universe; in \S~5 we summarize and discuss our findings.

Throughout this work we adopt the standard flat concordance cosmology (Planck Collaboration XIII 2016) with round parameter values: matter density $\Omega_M = 0.32$, baryon density $\Omega_b = 0.05$, Hubble constant $H_0 = 100\, h$ km s$^{-1}$ Mpc$^{-1}$ with $h = 0.67$, and mass variance $\sigma_8 = 0.83$ on a scale of $8\, h^{-1}$ Mpc. Stellar masses and luminosities (or SFRs) of galaxies are evaluated assuming the Chabrier's (2003) initial mass function (IMF).

\section{Cosmic star formation history}\label{sec|SFRfunc}

Our starting point is the global SFR function ${\rm d}N/{\rm d}\log \dot M_\star$, namely the number density of galaxies per logarithmic bin of SFR $[\log \dot M_\star,\log\dot M_\star+{\rm d}\log\dot M_\star]$ at given redshift $z$. This has been accurately determined by Mancuso et al. (2016), to which we defer the reader for details, recalling here only some basic aspects.

\begin{figure*}
\epsscale{0.95}\plotone{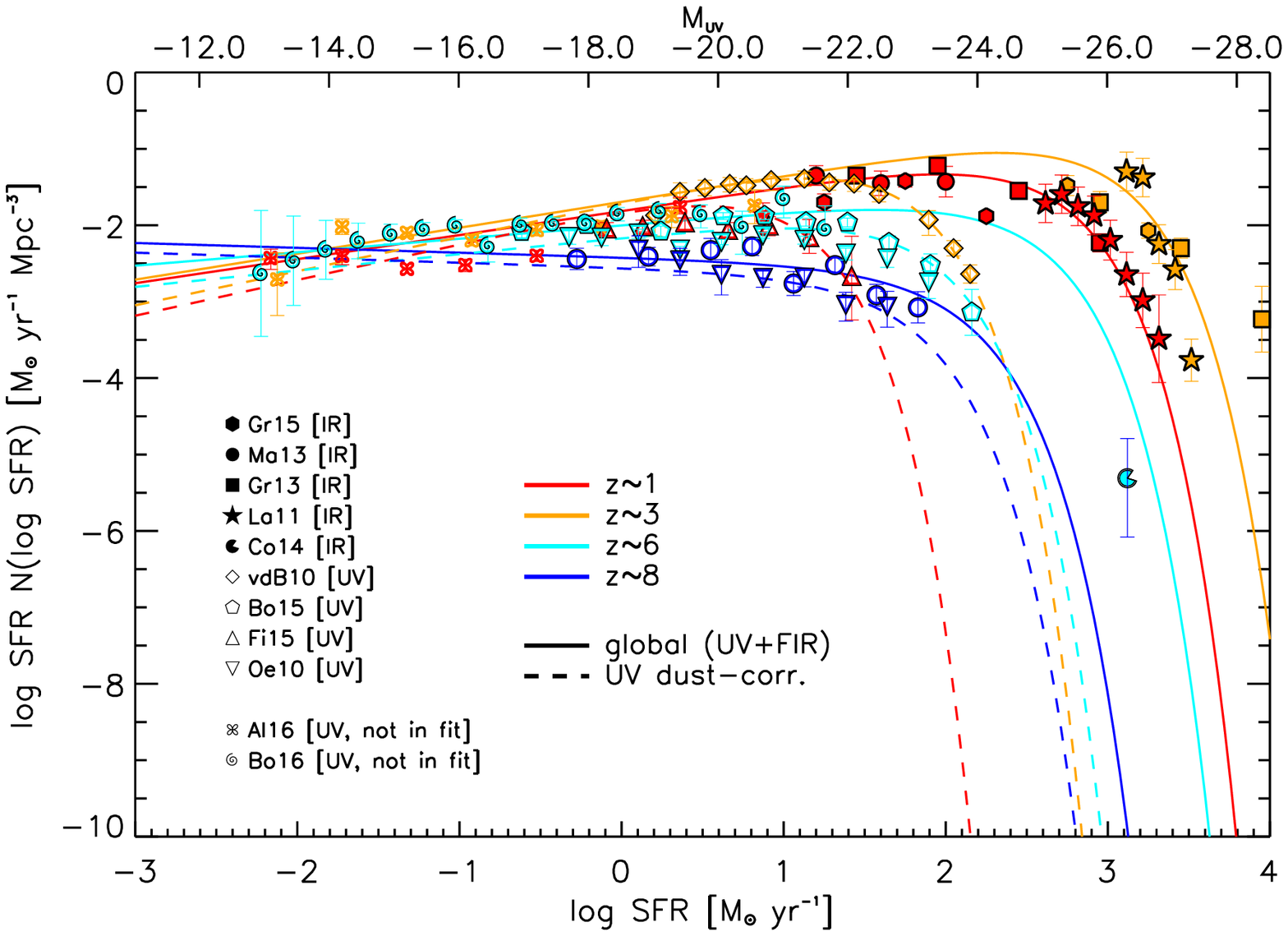}\epsscale{1.1}\plotone{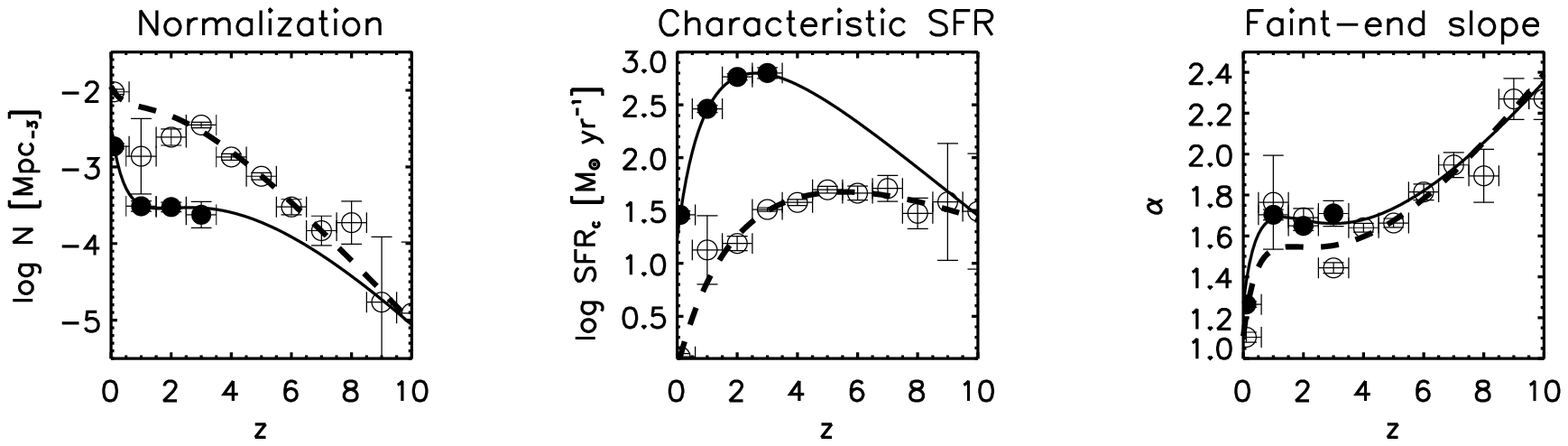}\caption{Top panel: SFR functions at different redshifts $z\sim 1$ (red lines), $3$ (orange), $6$ (cyan), and $10$ (blue), weighted by one power of the SFR; bottom axis refer to the SFR and top axis to the corresponding (intrinsic) UV magnitude. Solid lines show the global SFR functions proposed by Mancuso et al. (2016) from UV+far-IR data, while dashed lines show the SFR functions inferred only from UV data (dust corrected basing on the UV slope), see \S~\ref{sec|SFRfunc}. UV data (empty symbols) are from  van der Burg et al. (2010; diamonds), Bouwens et al. (2015; pentagons), Finkelstein et al. (2015; reverse triangles), Oesch et al. (2010; triangles), Alavi et al. (2016; clovers, not considered in the fit), and Bouwens et al. (2016b; spirals, not considered in the fit); IR data (filled symbols) are from Gruppioni et al. (2015; hexagons), Magnelli et al. (2013; circles), Gruppioni et al. (2013; squares), Lapi et al. (2011; stars), and Cooray et al. (2014; pacman). Bottom panels: parameters of the Schechter fits to the SFR functions; circles (empty for UV and filled for UV+far-IR) show the values at the specific $z$ where the fit to the data has been performed, while lines (dashed for UV and solid for UV+far-IR) are the corresponding polynomial rendition for the redshift evolution of each parameter (see \S~\ref{sec|SFRfunc} for details).}\label{fig|SFR_func}
\end{figure*}

In a nutshell, the SFR function has been built up by exploiting the most recent determinations of the luminosity functions at different redshifts down to $M_{\rm UV}\approx -17$ from far-IR and UV data (symbols in top panel of  Fig.~\ref{fig|SFR_func}). UV data have been dust-corrected according to the local empirical relation between the UV slope $\beta_{\rm UV}$ and the IR-to-UV luminosity ratio IRX (see Meurer et al. 1999), that is also routinely exploited for high-redshift galaxies (see Bouwens et al. 2009, 2015, 2016a,b). For the sake of simplicity, here we adopt a Meurer/Calzetti prescription, but we stress that the determination of the SFR functions is only marginally affected by choosing a different extinction law, like for example the Small Magellanic Cloud (SMC) one. Note that for violently star forming galaxies with intrinsic SFR $\dot M_\star\ga 30\, M_\odot$ yr$^{-1}$ the UV data, even when dust corrected via the UV slope-IRX relationship, strongly underestimate the intrinsic SFR, which is instead well probed by far-IR data. This is because high SFRs occur primarily within heavily dust-enshrouded molecular clouds, while the UV slope mainly reflects the emission from stars obscured by the diffuse cirrus dust component (Silva et al. 1998; Efstathiou et al. 2000; Efstathiou \& Rowan-Robinson 2003; Coppin et al. 2015; Reddy et al. 2015; Mancuso et al. 2016). On the other hand, at low SFR $\dot M_\star\la 10\, M_\odot$ yr$^{-1}$ the dust-corrected UV data efficiently probe the intrinsic SFR.

The luminosity $L$ has been converted into the SFR $\dot M_\star$ using $\log {\dot M_\star/ M_\odot~{\rm yr}^{-1}} \approx -9.8+\log {L/ L_\odot}$, a good approximation both for far-IR and (intrinsic) UV luminosities, as expected on energy conservation arguments, under the assumption of a Chabrier's IMF. Note that actually this conversion factor depends on the star formation history, and specifically on duration and age of the burst (see Efstathiou et al. 2000; Bressan et al. 2002); the standard value adopted here is the average for a continuous star formation over $100$ Myr, the age at which $90\%$ of emission has been contributed (see Kennicutt \& Evans 2012, their Table 1).

A smooth analytic representation of the SFR function has been determined in terms of the standard Schechter shape
\begin{equation}
{{\rm d}N\over {\rm d}\log\dot M_\star}(\dot M_\star,z) = \mathcal{N}(z)\, \left[\dot M_\star\over\dot M_{\star, c}(z)\right]^{1-\alpha(z)}\,e^{-\dot M_\star/\dot M_{\star, c}(z)}~,
\end{equation}
characterized at any given redshift $z$ by three parameters, namely, the normalization $\mathcal{N}$, the characteristic SFR $\dot M_{\star, c}$ and the faint end slope $\alpha$. We determine the values of the three Schechter parameters over the range $z\sim 0-10$ in unitary redshift bins $\Delta z\approx 1$ by performing an educated fit to the data. Specifically, for redshift $z\la 3$ UV data are fitted for SFRs $\dot M_\star\la 30\, M_{\odot}$ yr$^{-1}$ since in this range dust-corrections based on the $\beta_{\rm UV}$ are reliable, while far-IR data are fitted for SFRs $\dot M_\star\ga 10^2\, M_{\odot}$ yr$^{-1}$ since in this range dust emission is largely dominated by molecular clouds and reflects the ongoing SFR. On the other hand, for $z\ga 10$ the (dust-corrected) UV data are considered by themselves reliable estimators of the global SFR functions, since the amount of dust in a star-forming galaxy is expected to be rather small for an age of the Universe shorter than $5\times 10^8$ yr. In the bottom panels of Fig.~\ref{fig|SFR_func} we report as circles with error bars the values and uncertainties of the Schechter parameters at the specific redshifts where the fit to the data has been performed; empty symbols refer to the parameters for the (dust-corrected) UV-inferred SFR functions while filled symbols to the parameters for the global (UV+far-IR) SFR functions.

To obtain a smooth yet accurate representation of the SFR functions at any redshift, we find it necessary to (minimally) describe the redshift evolution for each parameter $p(z)$ of the Schechter function as third-order polynomial in log-redshift $p(z)=p_0+p_1\, \xi+p_2\,\xi^2+p_3\,\xi^3$, with $\xi=\log(1+z)$. The values of the parameters $\left\{p_i\right\}$
are reported in Table~1, and the polynomial fits to the redshift evolution of the Schechter parameters are shown as lines in the bottom panels of Fig.~\ref{fig|SFR_func}; dashed lines refer to the (dust-corrected) UV-inferred SFR functions while solid line to the global (UV+far-IR) SFR functions. The behavior of the normalization $\mathcal{N}$ and of the characteristic SFR $\dot M_{\star, c}$ highlight that UV surveys tend to pick up many galaxies with low SFR, while far-IR surveys also select less numerous galaxies with higher SFR; this mirrors the fact that high SFRs are usually associated to large dust abundance. The evolution with redshift of these parameters shows that most of the SFR occurs in dusty galaxies around redshift $z\approx 2$ (cf. also Fig.~\ref{fig|SFR_cosm}); on the other hand, toward high redshift $z\ga 6$ the dust content progressively decreases and UV surveys become more effective in selecting the typical population of star forming galaxies. Note that for the purpose of this paper, the high-redshift evolution is most relevant.

The resulting SFR functions (weighted by one power of the SFR) for representative redshifts $z\approx 1$ (red line), $3$ (orange), $6$ (cyan), $8$ (blue) are illustrated in Fig.~\ref{fig|SFR_func}. The top axis has been labeled using the relation $M_{\rm UV} \approx 18.5-2.5\, \log {\rm SFR} [M_{\odot}$ yr$^{-1}]$ between the intrinsic UV magnitude and the SFR; at the faint end for $M_{\rm UV}\la -20$ where dust extinction is negligible, this also provides an estimate of the observed UV magnitude. All in all, at $z\ga 4$ our global estimate implies a significant number density of dusty starforming galaxies with SFR $\dot M_\star\ga 10^2\, M_{\odot}$ yr$^{-1}$, currently missed by UV data, even when corrected via the standard UV slope. To highlight more clearly this point, we also report in Fig.~\ref{fig|SFR_func} the SFR function that would have been inferred basing solely on the (dust-corrected) UV data.

In Mancuso et al. (2016; 2017) we have validated the global SFR functions against independent datasets, including galaxy number counts at significative submm/far-IR wavelengths, redshift distributions of gravitationally lensed galaxies, cosmic infrared background, galaxy stellar mass function via the continuity equation, main sequence of starforming galaxies, and even associated AGN statistics. In particular, the analysis of the
main sequence for high-redshift galaxies and AGNs presented in Mancuso et al. (2017) highlights that the current data can be consistently interpreted in terms of an in situ coevolution scenario for star formation and black hole accretion, envisaging these as local, time coordinated processes.

In the present paper, when dealing with the reionization history of the Universe or with the estimates of the cosmic SFR density from GRB rates, the behavior of the SFR function at the very faint end down to $M_{\rm UV}\approx -12$ (corresponding to SFR $\dot M_\star\approx$ a few $10^{-3}\, M_\odot$ yr$^{-1}$) will become relevant. This actually requires to extrapolate our SFR functions well beyond the magnitudes limit $M_{\rm UV}\approx -17$ (corresponding to SFR $\dot M_\star\approx$ a few $10^{-1}\, M_\odot$ yr$^{-1}$) currently accessible to blank field UV surveys, that we have exploited in our fitting procedure (see above).

One may wonder whether these extrapolations of the SFR functions are reasonable, and in particular whether the faint end slope keeps steep values $\alpha\la 2$ as inferred for $M_{\rm UV}\ga -17$. Actually, at redshifts $z\la 6$ the faint end of the luminosity/SFR functions has been recently explored, though still with large uncertainties, even down to $M_{\rm UV} \approx -13$ via gravitational lensing by foreground galaxy clusters (see Alavi et al. 2014, 2016; Livermore et al. 2016; Bouwens et al. 2016). We have reported (but not used in the fit because of the still large systematic uncertainties) these data in Fig.~\ref{fig|SFR_func} to highlight that they are indeed consistent with the extrapolation of our SFR functions; specifically, the faint portion of the SFR function is seen to keep rising steeply with $\alpha\la 2$ out to $z\la 6$.

In moving toward yet higher redshifts $z\sim 8$, our SFR function determination based on the current UV data down to $M_{\rm UV}\approx -17$ feature a faint end slope steepening to values $\alpha\ga 2$. This is somewhat theoretically expected, since it reflects the steepening in the underlying mass function of dark matter halos, given that the slope of the SFR vs. halo mass relationship between $z\approx 3$ and $z\approx 6-8$ is not appreciably different (see Aversa et al. 2015; also cf. Fig.~\ref{fig|abundmatch}).

\subsection{Cosmic star formation rate density}\label{sec|SFR_cosm}

We now turn to exploit our SFR functions to compute, and illustrate in Fig.~\ref{fig|SFR_cosm}, the cosmic SFR density
\begin{equation}\label{eq|SFR_func}
\rho_{\rm SFR}(z) = \int_{\dot M_\star^{\rm min}}^\infty{\rm d}\log \dot M_{\star}\, {{\rm d}N\over {\rm d}\log \dot M_\star}\, \dot M_\star~;
\end{equation}
here $\dot M_\star^{\rm min}$ is a minimum SFR limit, relevant because of the steepness $\alpha>1$ of the SFR functions at the faint end.

We start by considering the UV-inferred SFR functions (dust corrected via the UV slope) integrated down to an UV magnitude $M_{\rm UV}^{\rm min}\approx -17$ (corresponding to a minimum SFR $\dot M_\star^{\rm min}\approx$ a few $10^{-1}\, M_\odot$ yr$^{-1}$) that matches the observational limit of current blank-field UV surveys. The outcome (blue dashed line) is in good agreement with the dust corrected data by Bouwens et al. (2015; cyan squares) at $z\ga 4$ and by Schiminovich et al. (2005; cyan shaded area) at $z\la 4$.

On the other hand, the cosmic SFR density from (dust-corrected) UV data is inconsistent with other datasets both at low and high redshift. Specifically, at redshift $z\la 4$ it falls short with respect to the multiwavelength determination by Hopkins \& Beacom (2006; orange shaded area) based on UV/optical, radio, H$\alpha$ and mid-IR $24\, \mu$m data, and to the far-IR measurements from \textsl{Herschel} by Magnelli et al. (2013) and Gruppioni et al. (2013; red shaded area). At redshift $z\ga 4$ it underestimates the determinations based on stacking of far-IR data from \textsl{Herschel} by Rowan-Robinson et al. (2016; red dots), and the determination based on long GRB rates from \textsl{Swift} by Kistler et al. (2009; 2013). This mostly reflects the fact, already mentioned above, that the UV-inferred SFR functions (even corrected for dust extinction by the UV slope) appreciably underestimate the number density of dusty galaxies with $\dot M_\star\ga 30\, M_\odot$ yr$^{-1}$.

Thus we compute the cosmic SFR density exploiting our global (far-IR+UV) SFR function down to the same magnitude limit $M_{\rm UV}^{\rm min}\approx -17$. The outcome (red dashed line) is found to be in good agreement both with the Hopkins \& Beacom (2006) and Gruppioni et al. (2013) determinations at $z\la 4$ and with the stacked far-IR data by Rowan-Robinson et al. (2016) at $z\ga 4$. However, at $z\ga 6$ the results is only marginally consistent with the estimates from GRB rates; this hints towards a substantial contribution to the cosmic SFR density from faint GRB hosts with $M_{\rm UV}\ga -17$.

In particular, we show the outcome (solid line) when integrating our global SFR functions down to $M_{\rm UV}^{\rm lim}\approx -13$ (corresponding to a minimum SFR $\dot M_\star^{\rm min}\approx 10^{-2}\, M_\odot$ yr$^{-1}$); as we shall discuss in the next Section, this value is also indicated by the recent data on the reionization history of the Universe in terms of electron scattering optical depth $\tau_{\rm es}\approx 0.058$ as measured by \textsl{Planck} Collaboration et al. XLVII (2016). Our result compares fairly well with the GRB data at $z\ga 6$ but appears to overestimate those around $z\sim 4-5$. Actually this is due to the well established fact that long GRBs tend to occur mostly in low-metallicity environments. Thus we add this further constraint, weighting the integrand in Eq.~(\ref{eq|SFR_func}) with the time spent by an individual galaxy below a given critical metallicity ceiling $Z_{\rm crit}$. To compute the latter, we have exploited a standard code of chemical evolution (see Feldmann et al. 2015) that reproduces the locally observed mass-metallicity relation (e.g., Zahid et al. 2014). We find, in pleasing agreement with metallicity estimates at $z\la 6$ from GRB afterglow spectra (e.g., Salvaterra et al. 2013; Perley et al. 2016) and with GRB formation models based on binary evolution (e.g., Fryer \& Heger 2005; Podsiadlowski et al. 2010; see also review by Kumar \& Zhang 2015), that a value $Z_{\rm crit}\approx Z_\odot/2$ reproduces fairly well the GRB-inferred SFR density (green solid line) over the whole redshift range $z\ga 4$.

We stress that further decreasing the limiting magnitude below $M_{\rm UV}^{\rm lim}\approx -13$ does not change substantially the cosmic SFR density for $z\la 8$; e.g., the dot-dashed red line shows the outcome when adopting $M_{\rm UV}^{\rm lim}\approx -12$ (corresponding to a minimum SFR $\dot M_\star^{\rm min}\approx$ a few $10^{-1}\, M_\odot$ yr$^{-1}$), that will turn out to be the faintest value allowed by the \textsl{Planck} data on cosmic reionization and by the missing satellite issue (cf. \S~\ref{sec|reion} and \ref{sec|sats}).

\section{Reionization history}\label{sec|reion}

Another important observational channel, plainly connected with the cosmic SFR density, is the reionization of the Universe as probed by the electron scattering optical depth $\tau_{\rm es}$. This is because the cosmic ionization rate $\dot N_{\rm ion}$ is just proportional to the cosmic SFR density
\begin{equation}
\dot N_{\rm ion}\approx f_{\rm esc}\, k_{\rm ion}\, \rho_{\rm SFR}~;
\end{equation}
here $k_{\rm ion}\approx 4\times 10^{53}$ is the number of ionizing photons s$^{-1}$ $(M_\odot/\rm yr)^{-1}$, with the quoted value appropriate for a Chabrier IMF, and $f_{\rm esc}\la 0.2$ is the average escape fraction for ionizing photons from the interstellar medium of high-redshift galaxies (see Mao et al. 2007; Dunlop et al. 2013; Robertson et al. 2015). Note that two other parameters implicitly entering in the expression $\rho_{\rm SFR}$ are the minimum UV limiting magnitude, and the faint-end slope of the SFR functions $\alpha$ discussed at the end of \S~\ref{eq|SFR_func}.

\begin{figure*}
\epsscale{1}\plotone{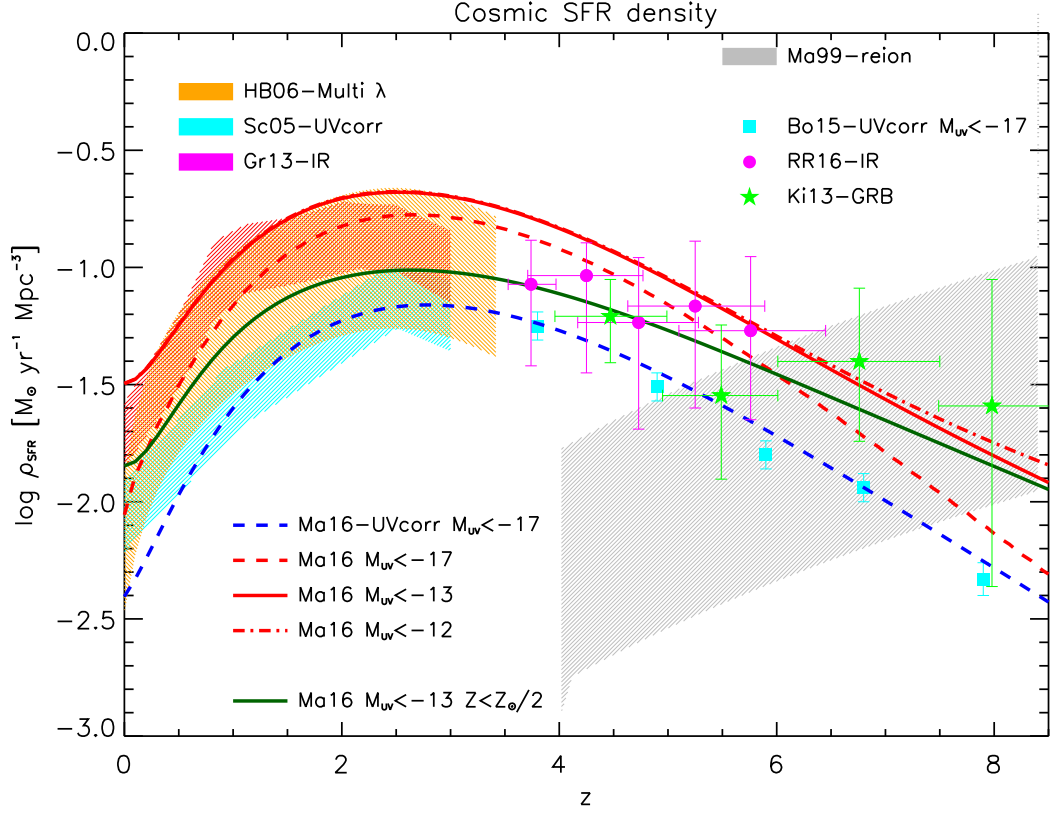}\caption{The cosmic star formation rate density as a function of redshift. Blue dashed line shows the outcome for
the dust-corrected UV-inferred SFR functions integrated down to a
limiting magnitude $M_{\rm UV}^{\rm lim}\approx -17$. Red lines illustrate the outcomes from our global (UV+IR) SFR functions integrated down to UV limiting magnitudes $M_{\rm UV}^{\rm lim}\approx -17$ (dashed), $-13$ (solid), and $-12$ (dot-dashed). The green solid line is again for $M_{\rm UV}^{\rm lim}\approx -13$ but with a ceiling in metallicity set to $Z<Z_\odot/2$. At $z\la 4$ data are from: (dust-corrected) UV observations by Schiminovich et al. (2005; cyan shaded area); far-IR observations by Gruppioni et al. (2013; red shaded area); multiwavelength determination including UV, radio, H$\alpha$ and mid-IR $24\, \mu$m data collected by Hopkins \& Beacom (2006; orange shaded area). At higher redshift $z\ga 4$, we report the estimate of the SFR density inferred from: (dust-corrected) UV data by Bouwens et al. (2015; cyan squares), stacked IR data by Rowan-Robinson et al. (2016; magenta circles), long GRB rates by Kistler et al. (2009, 2013; green stars). The grey shaded area is the estimate of the critical SFR density for cosmic reionization from Madau et al. (1999) for the range of values $C_{\rm HII}/f_{\rm esc}\sim 10$ (upper envelope) to $100$ (lower envelope) in the ratio between the clumping factor of ionized hydrogen and escape fraction of ionizing photons from high-$z$ starforming galaxies.}\label{fig|SFR_cosm}
\end{figure*}

The ionization rate rules the standard evolution equation of the HII ionizing fraction
\begin{equation}
\dot Q_{\rm HII} = {\dot N_{\rm ion}\over \bar n_{\rm H}}-{Q_{\rm HII}\over
t_{\rm rec}}
\end{equation}
that takes into account the competition between ionization and recombination
processes (see Madau et al. 1999; Ferrara \& Pandolfi 2015). In the above equation $\bar n_{\rm H}\approx 2\times 10^{-7}\, (\Omega_b h^2/0.022)$ cm$^{-3}$ is the mean comoving hydrogen number density. In addition, the recombination timescale reads $t_{\rm rec}\approx 3.2$ Gyr $[(1+z)/7]^{-3}\, C_{\rm HII}^{-1}$, where the case B coefficient for an IGM temperature of $2\times 10^4$ K has been used; this timescale crucially depends on the clumping factor of the ionized hydrogen, for which a fiducial value $C_{\rm HII}\approx 3$ is usually adopted (see Pawlik et al. 2013). A first glimpse on the level of cosmic SFR density required to balance recombination is provided by Madau et al. (1999), and it is reported as a grey shaded area in Fig.~\ref{fig|SFR_cosm} for different values of the ratio $C_{\rm HII}/f_{\rm esc}\sim 10-100$.

The electron scattering optical depth is then obtained by integrating the ionized fraction over redshift
\begin{equation}
\tau_{\rm es}(z) = c\, \sigma_{\rm T}\,\bar n_{\rm H}\int^z{\rm d}z'\,f_e\,Q_{\rm HII}(z')
(1+z')^2 \, H^{-1}(z')~;
\end{equation}
here $H(z)=H_0\,[\Omega_M\,(1+z)^3+1-\Omega_M]^{1/2}$ is the Hubble
parameter, $c$ is the speed of light, $\sigma_{\rm T}$ the Thomson cross
section and $f_e$ the number of free-electron (assuming double
Helium ionization at $z\lesssim 4$.).

\begin{figure*}
\epsscale{1}\plotone{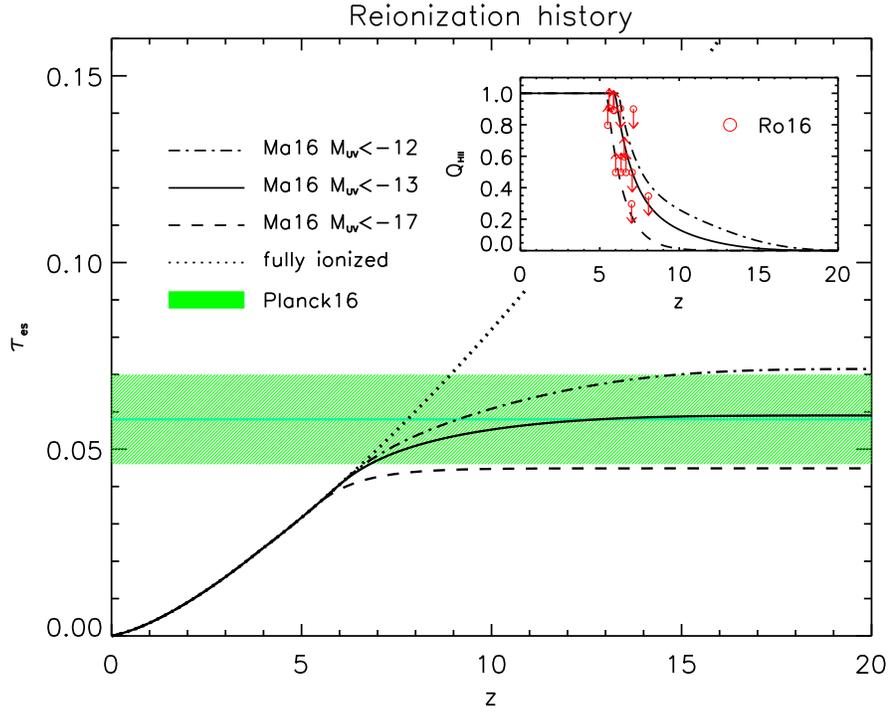}\caption{Reionization history of the Universe, in terms of the redshift evolution of the optical depth $\tau_{\rm es}$ for electron scattering. Black lines illustrate the outcomes from our SFR functions integrated down to UV magnitude limits $M_{\rm UV}^{\rm lim}\approx -13$ (solid) $-17$ (dashed) and $-12$ (dot-dashed) when adopting an escape fraction of ionizing photons $f_{\rm esc}\approx 0.1$, see \S~\ref{sec|reion}. For reference, the black dotted line refers to a fully ionized Universe up to redshift $z$. The green shaded area shows the measurement (with 1$\sigma$ uncertainty region) from the \textsl{Planck} Collaboration et al. XLVII (2016). In the inset, the corresponding evolution of the ionized fraction $Q_{\rm HII}$ is plotted, together with upper and lower limits from various observations as collected by Robertson et al. (2015; empty circles).}\label{fig|reion}
\end{figure*}

In Fig.~\ref{fig|reion} we illustrate the cosmic reionization history computed from our global SFR function integrated down to different $M_{\rm UV}^{\rm lim}$, on assuming a conservative value $f_{\rm esc}\approx 0.1$ for the escape fraction of ionizing photons. When adopting $M_{\rm UV}^{\rm lim}\approx -13$, the outcome (black solid line) agrees with the value of the optical depth for electron scattering $\tau_{\rm }\approx 0.058$ recently measured by the \textsl{Planck} mission. For reference, the dotted line represents the optical depth expected in a fully ionized Universe up to redshift $z$; this is to show that the bulk of the reionization process occurred at $z \sim 8-9$ and was almost completed at $z \sim 6$ (see Schultz et al. 2014). Note that from this perspective, the detailed behavior of the SFR functions at $z\la 6$ and at $z\ga 10$, and the related ionizing background, are only marginally relevant. Remarkably, the evolution of the ionized fraction $Q_{\rm HII}$ illustrated in the inset is fully consistent with upper and lower limits from the plethora of independent observations collected by Robertson et al. (2015).

When adopting $M_{\rm UV}^{\rm lim}\approx -17$, that corresponds to the observational limits of current blank-field UV surveys at $z\ga 6$ (cf. Fig.\ref{fig|SFR_cosm}), the outcome on the optical depth (black dashed line) touches the lower boundary of the 1$\sigma$ region allowed by \textsl{Planck} data, and the evolution of the $Q_{\rm HII}$ parameter is inconsistent with the aforementioned observational limits. At the other end, going much beyond $M_{\rm UV}^{\rm lim}\approx -13$ is not allowed, since already for $M_{\rm UV}^{\rm lim}\approx -12$ the resulting optical depth (black dot-dashed line) touches the upper boundary of the 1$\sigma$ region from \textsl{Planck} data.

Note that these outcomes suffer to some extent of parameter degeneracy, as highlighted by the expression
\begin{equation}
f_{\rm esc}\, k_{\rm ion}\,C_{\rm HII}^{-0.3}\, \Gamma[2-\alpha;\dot M_\star^{\rm lim}/\dot M_{\star,c}]\approx {\rm const}
\end{equation}
where $\Gamma[a;z]\equiv \int_z^\infty{\rm
d}x~x^{a-1}\,e^{-x}$ is the incomplete $\Gamma-$function; the parameters involved are the escape fraction $f_{\rm esc}$, the ionizing rate per unit SFR $k_{\rm ion}$ associated mainly to the IMF, the clumping factor $C_{\rm HII}$, the faint-end slope of the SFR function $\alpha$, and the UV magnitude limit $M_{\rm UV}^{\rm lim}$. The strongest dependencies are on $f_{\rm esc}$ and on the limiting magnitude $M_{\rm UV}^{\rm lim}$. For example, to reproduce the \textsl{Planck} best estimate $\tau_{\rm es}\approx 0.058$ for $M_{\rm UV}^{\rm lim}\approx -17$, it would be necessary to force $f_{\rm esc}$ to implausible values $\ga 0.2$; at the other end, setting $M_{\rm UV}^{\rm lim}\approx -12$ would require $f_{\rm esc}\la 0.05$, but at the cost of worsening the agreement with the observational constraints on $Q_{\rm HII}$.

All in all, the adoption of $M_{\rm UV}^{\rm lim}\approx -13$ with a conceivable escape fraction of $f_{\rm esc}\approx 0.1$ constitutes the best compromise to reproduce the \textsl{Planck} constraints on the reionization history of the Universe, while still retaining consistency with the cosmic SFR density inferred from IR and GRB data at high $z\ga 6$ (see Fig.~ \ref{fig|SFR_cosm}).

\section{Missing satellites}\label{sec|sats}

We now aim at converting such constraints on the UV limiting magnitude (or equivalently on the minimum SFR) into bounds on masses of the host dark matter halos. To this purpose, we need an average statistical relationship between the SFR (or the UV magnitude) of a galaxy and its host halo mass; we obtain the latter via the standard abundance matching technique, i.e.,
by associating galaxies and halos with the same integrated number density (e.g., Vale \& Ostriker 2004; Shankar et al. 2006; Moster et al. 2013; Aversa et al. 2015).

\begin{figure*}
\epsscale{0.9}\plotone{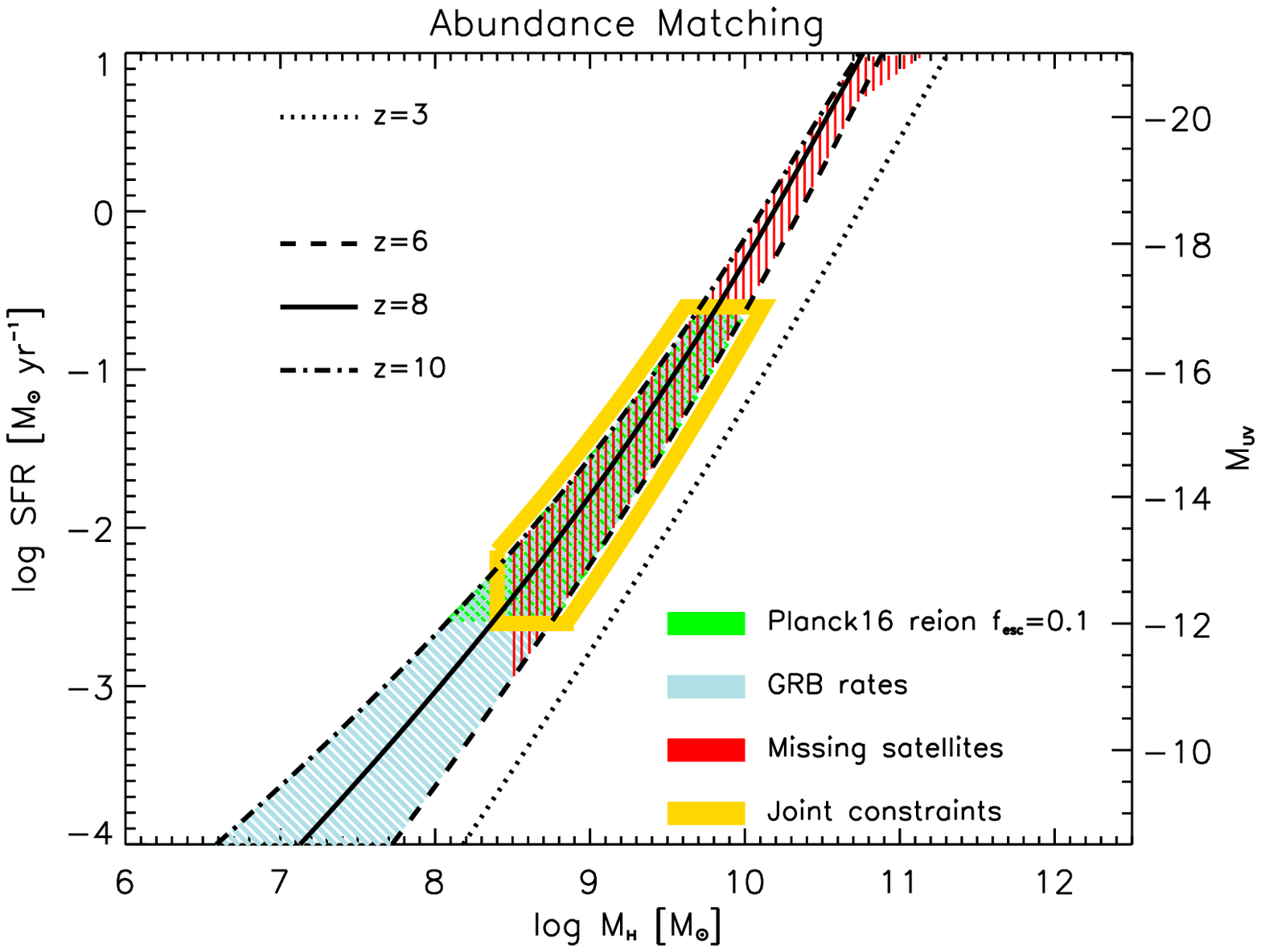}\\\epsscale{1.1}\plottwo{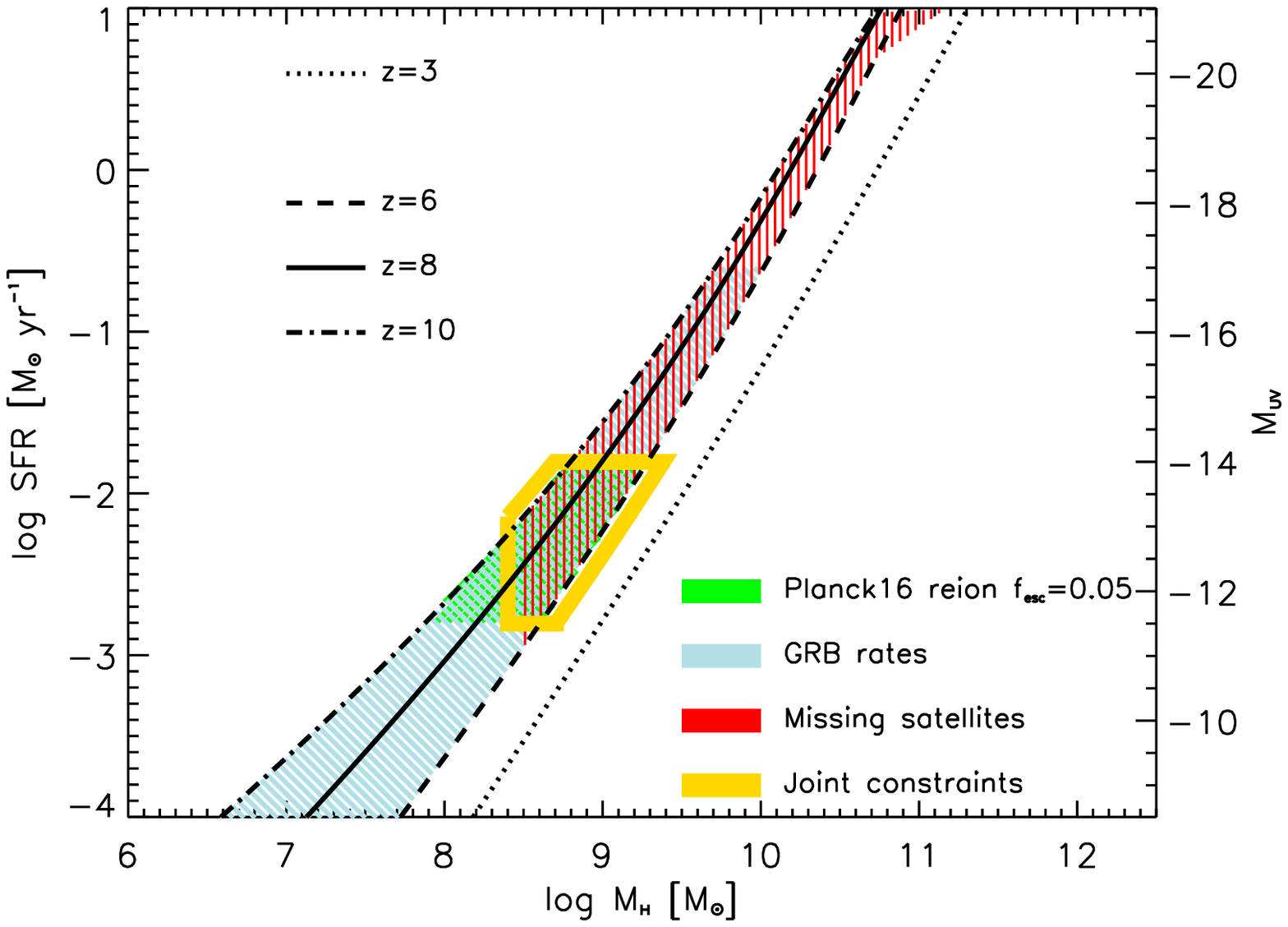}{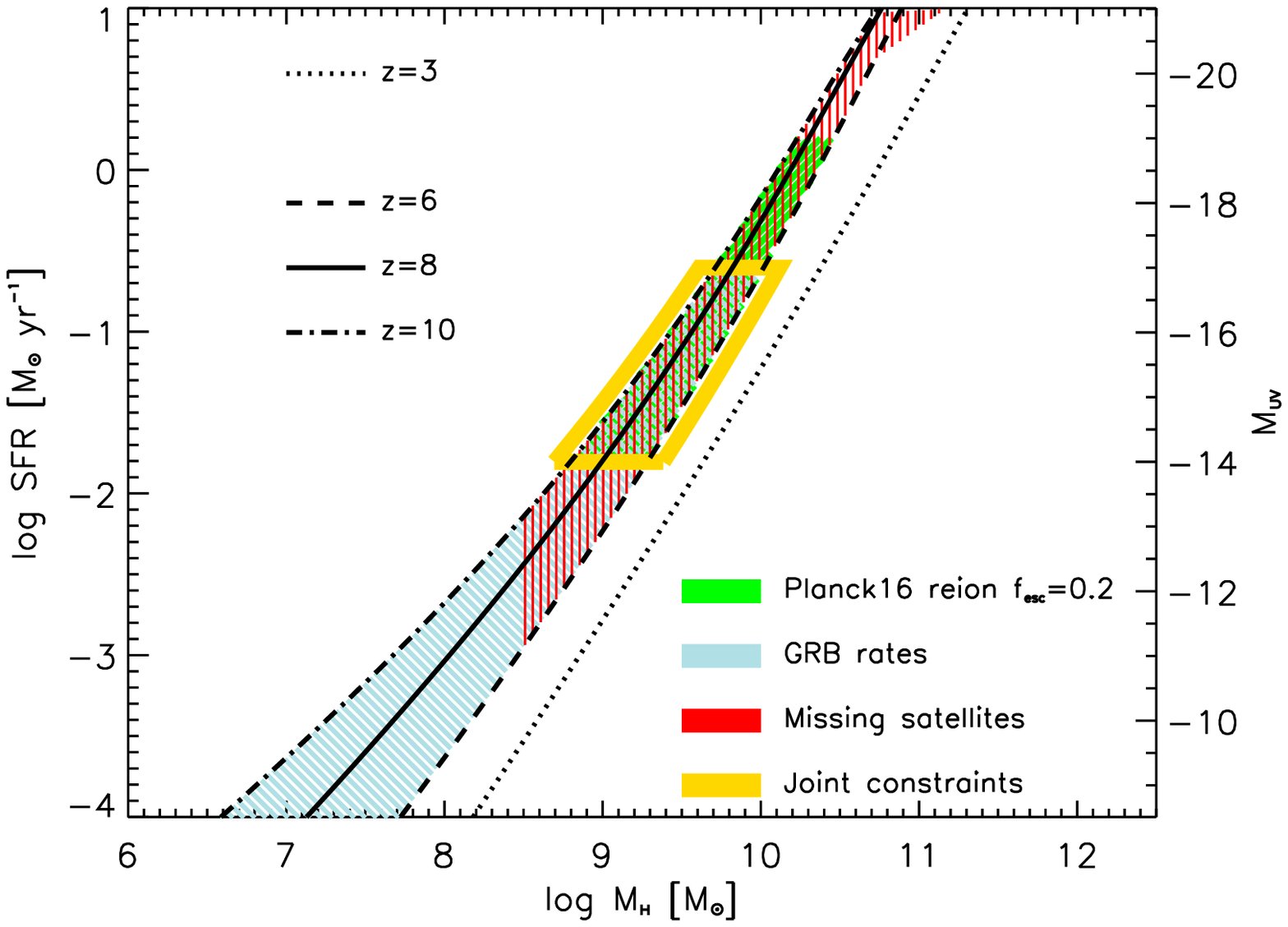}
\caption{The relationship between SFR (left axis) or UV magnitude (right axis) and halo mass at $z\sim 6$ (dashed line), $8$ (solid), and $10$ (dot-dashed), as derived via the abundance matching technique from our global SFR functions and the halo mass function; for reference, the relation at $z\sim 3$ (dotted) is also shown. The green hatched area marks the region of limiting magnitudes required to reproduce the Planck data on the electron scattering optical depth, with an escape fraction $f_{\rm esc}\approx 0.1$ (top panel), $0.05$ (bottom left) and $0.2$ (bottom right). The light blue hatched area refers to the corresponding constraints from GRB rates, independent of the escape fraction. The red hatched area marks the region where galaxy formation is allowed to occur from the missing satellite issue (see Boylan-Kolchin et al. 2014; Wetzel et al. 2016). The yellow contours highlight the joint constraints.}\label{fig|abundmatch}
\end{figure*}

The outcome for our global SFR function at three redshifts $z\sim 6$ (black dashed line), $8$ (solid), and $10$ (dotted) relevant for reionization is illustrated in Fig.~\ref{fig|abundmatch}. We find a relationship $\dot M_\star\propto M_{\rm H}^{1.5}$ between the SFR and host halo mass, which is remarkably similar to what has been derived at lower $z\la 6$ (e.g., Aversa et al. 2015); the slope is close to that expected for galaxies where the SFR is regulated by the balance between cooling and energy feedback from supernova explosions or stellar winds (see Shankar et al. 2006; Finlator et al. 2011).

We stress once more that the current UV data from blank-field surveys  at $z\ga 6$ probe the SFR functions down to $M_{\rm UV}\approx -17$, so actually most of these relations are obtained by extrapolation. However, we also report the relation at $z\approx 3$ where the SFR function has been probed down to $M_{\rm UV}\approx -12$. The relationships at $z\la 3$ and at $z\ga 6$ have approximately the same slope, hinting toward a common \emph{in situ} nature of the star formation process within galaxies (see Mancuso et al. 2016; 2017). The offset in normalization is theoretically well understood in terms of an increased efficiency of star formation at high redshift, because of shorter cooling times induced by the denser cosmic environment (cf. Lapi et al. 2011; 2014).

We also report constraints on the limiting $M_{\rm UV}^{\rm lim}$ given by the analysis of the cosmic SFR inferred from GRBs (light blue hatched area), and those from \textsl{Planck} data on $\tau_{\rm es}$ (green area). We stress that, while the GRB constraints are much looser, the one on $\tau_{\rm es}$ depends critically on the assumed value $f_{\rm esc}\approx 0.1$ for the escape fraction (cf. \S~\ref{sec|reion}). In the bottom panels, the reader may appreciate how changing $f_{\rm esc}$ to $0.05$ or to $0.2$ alters the \textsl{Planck} constraints.

Then we add the independent constraint from the missing satellite problem issue (red area). As shown by numerical simulations (Boylan-Kolchin et al. 2014; Wetzel et al. 2016), astrophysical processes must intervene to severely limit or even suppress galaxy formation in halos with low masses $M_{\rm H}\la$ few $\times 10^8\,M_\odot$, because a substantial number of such small halos would otherwise survive down to the present time as bound satellites of Milky Way-sized galaxies, which are not observed. This is reinforced by the recent estimates of the halo masses from kinematics of local dwarf spirals: the objects with the smallest stellar masses $M_\star\approx 10^7\, M_\odot$ are found to reside in halos with masses always larger than a few $\times 10^8\,M_\odot$ (see Karukes et al. 2017).

Physical processes suppressing galaxy formation in such small halos may include an increase of supernova feedback efficiency, or radiative/chemical feedback from the diffuse UV background, or more complex phenomena (see Efstathiou et al. 1992; Sobacchi et al. 2013; Cai et al. 2014). Note that an alternative, though more exotic, possibility is constituted by warm dark matter with particle masses $\la 3$ keV (Viel et al. 2013; Lapi \& Danese 2015): in such frameworks halo masses below a few $\times 10^8\,M_\odot$ may not exist at all due to the truncation in the primordial power spectrum, and even cooling/star formation processes may be less efficient due to the lack of substructures (e.g., Pontzen \& Governato 2014).

We remark that the combination of the astrophysical/cosmological constraints from the satellite issue, GRB rates, and reionization (highlighted in Fig.~\ref{fig|abundmatch} by the yellow contour) concurrently support a limiting UV magnitude around $M_{\rm UV}^{\rm lim}\approx -13$ and a plausible value of the escape fraction around $f_{\rm esc}\approx 0.1$, producing a reionization redshift $z\approx 8-9$ and agreement with the constraints on the evolution of the HII ionizing fraction $Q_{\rm HII}$, cf. inset of Fig.~\ref{fig|reion}.

\begin{figure*}
\epsscale{1}\plotone{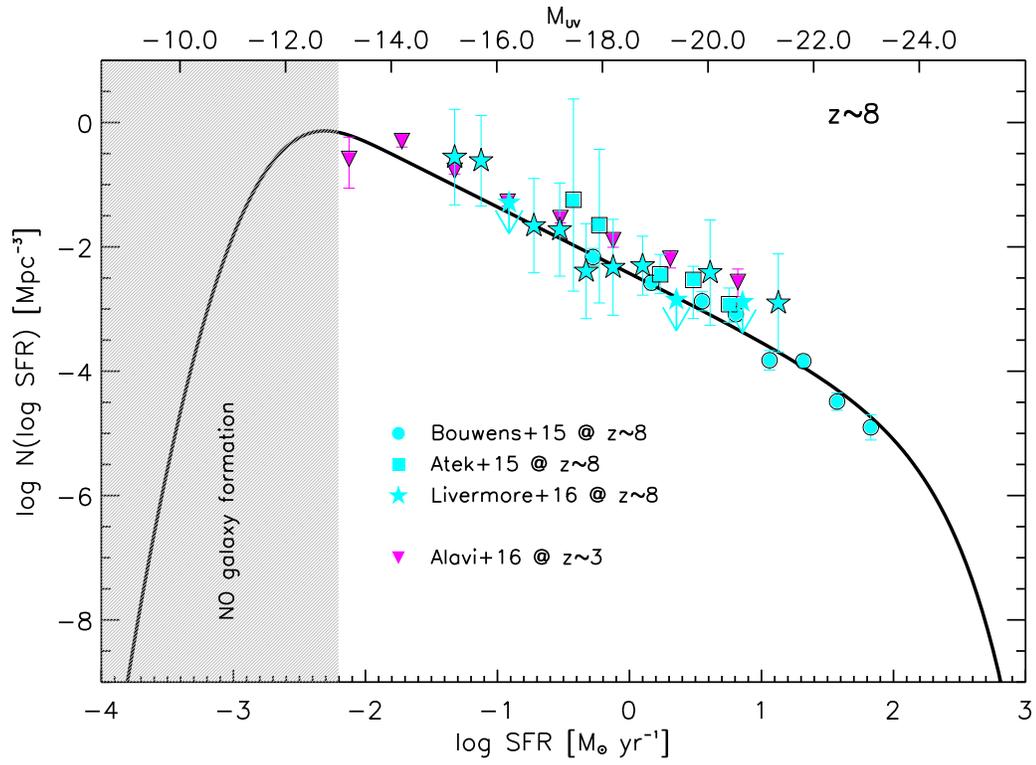}\caption{Solid line illustrates our prediction concerning the very faint end of the SFR function at $z\sim 8$; bottom axis refers to the SFR, while top axis to the corresponding intrinsic UV magnitude. Data at $z\sim 8$ are from Bouwens et al. (2015; cyan circles), Atek et al. (2015; cyan squares), and Livermore et al. (2016; cyan stars); for reference we also report the estimate at $z\sim 3$ from Alavi et al. (2016; magenta inverse triangles) that extends down to $M_{\rm UV}\approx -12$. The grey shaded area marks the region where galaxy formation must be inefficient, as inferred from the joint constraints from GRB rates, cosmic reionization, and missing satellites.}\label{fig|SFR_func_prediction}
\end{figure*}

In Fig.~\ref{fig|SFR_func_prediction} we provide a specific prediction concerning the very faint end of the SFR function at $z\sim 8$, based on the combined constraints from reionization, GRB rates and missing satellites (grey shaded area). These concurrently indicate the SFR function to downturn at around $M_{\rm UV}\approx -12$, not far from the magnitudes $M_{\rm UV}\approx -15$ that have been already observed by Atek et al. (2015) and Livermore et al. (2016) thanks to gravitational lensing effects. Remarkably, at $z\approx 3$ and $z\approx 6$ Alavi et al. (2016) and Bouwens et al. (2016) have pushed our knowledge of the SFR function down to $M_{\rm UV}\approx -13$, with no significant evidence for a downturn.

At $z\sim 8$ the relevant magnitude range is not accessible with current facilities but will be probed in the next future with the advent of the \textsl{James Webb Space Telescope}. The precise location and the detailed shape of the SFR function near the downturn would be extremely informative on the astrophysics of galaxy formation in small halos (e.g., Weisz et al. 2014).

\section{Summary}\label{sec|summary}

We have provided an holistic view of galaxy evolution at high redshift $z\ga 4$, by exploiting the constraints from various astrophysical and cosmological probes. These include the estimate of the cosmic SFR density from UV/IR surveys and long GRB rates, the cosmic reionization history after the latest \textsl{Planck} data, and the missing satellites issue.

Our arguments are model-independent, since they are based on the SFR functions at different redshifts $z\sim 0-10$ derived by Mancuso et al. (2016) via educated extrapolation of the latest UV/far-IR data from \textsl{HST}/\textsl{Herschel}. These SFR functions have already been extensively validated against a number of independent observables, including galaxy number counts at significative submm/far-IR wavelengths, redshift distributions of gravitationally lensed galaxies, cosmic infrared background, galaxy stellar mass function, main sequence of starforming galaxies, and associated AGN statistics. Our SFR functions imply at $z\ga 4$ a significant number density of dusty starforming galaxies with SFR $\dot M_\star\ga 10^2\, M_{\odot}$ yr$^{-1}$, currently missed by UV data even when corrected for dust extinction via the UV slope (cf. \S~\ref{sec|intro}). We stress that our model-independent SFR functions constitute a crucial test of galaxy formation, and specifically so for redshift $z\ga 2$ and large SFRs $\dot M_\star\ga 100\, M_\odot$ yr$^{-1}$, where most of the current numerical and semianalytic models suffer substantial problems (e.g., Gruppioni et al. 2015, their Fig.~3).

We have found that the cosmic SFR density inferred from our global SFR functions integrated down to an UV magnitude limit $M_{\rm UV}\la -13$ (corresponding to an SFR limit around $10^{-2}\, M_\odot$ yr$^{-1}$) is in excellent agreement with recent determinations from stacked IR data at redshift $z\ga 4$. Moreover, when taking into account a critical metallicity ceiling $Z\la Z_\odot/2$, it is also in accord with the estimates from long GRB rates (cf. \S~\ref{sec|SFR_cosm}). Note that such a value of the metallicity ceiling lends support to GRB formation models based on binary stellar evolution. We stress that such findings on the cosmic SFR and metallicity demonstrate the relevance of the issue concerning dust production mechanisms by type-II supernovae and AGB stars in high$-z$ galaxies (e.g., Dwek et al. 2015; Michalowski 2015; Mancuso et al. 2016; Rowan-Robinson et al. 2016). Large statistical samples of dusty star-forming galaxies at $z\ga 4$ will constitute key datasets for understanding the role of the physical
processes involved in dust formation and destruction.

We have been able to reproduce the recent measurements of the \textsl{Planck} mission on the electron scattering optical depth $\tau_{\rm es}\approx 0.058$, and to satisfy the constraints on the redshift evolution of the HII ionizing fraction $Q_{\rm II}$, by adopting a standard Chabrier IMF and a conceivable value of the average escape fraction for ionizing photons $f_{\rm esc}\approx 0.1$. At the same time we have shown that to retain consistency with the \textsl{Planck} data, the SFR functions at $z\ga 6$ cannot extend with steep slopes $\alpha\la 2$ much beyond $M_{\rm UV}\la -12$ (SFRs of a few $10^{-3}\, M_\odot$ yr$^{-1}$). Coordinated variations of the escape fraction and of the limiting magnitude can work comparably well in reproducing the estimate on $\tau_{\rm es}$ from \textsl{Planck}, but at the cost of worsening the agreement with the observational constraints on $Q_{\rm HII}$.

We have then connected via the abundance matching technique the constraints on the cosmic SFR density from IR/GRB data and on cosmic reionization to the missing satellite issue. All these independent probes concurrently imply galaxy formation to become inefficient (at least at $z\ga 6$) within dark matter halos of mass below a few $10^8\, M_\odot$, or correspondingly within galaxies with $M_{\rm UV}\la -12$ or SFRs of a few $10^{-3}\, M_\odot$ yr$^{-1}$. This inefficiency may be due to baryonic processes like increased supernova feedback efficiency,  radiative/chemical feedback from the diffuse UV background, or more complex phenomena. An alternative, though more exotic, possibility is constituted by the dark matter to be warm with particle masses $\la 3$ keV, implying that halo masses below a a few $\times 10^8\,M_\odot$ may be strongly suppressed in number due to the truncation in the primordial power spectrum by free-streaming, or that cooling/star formation processes may be less efficient in such small halos due to the lack of substructures.

Therefore, we have predicted a downturn of the galaxy luminosity function faintward of $M_{\rm UV}\la -12$, and stressed that its detailed shape, as plausibly probed in the next future by the \textsl{Hubble Frontier Fields} program or by the \textsl{JWST}, will be extremely informative on the astrophysics of galaxy formation in small halos, or even pivotal to probe the microscopic nature of the dark matter.

\begin{acknowledgements}
We thank the anonymous referee for constructive comments. We are grateful to A. Bressan, J. Beacom, E. Karukes, and P. Salucci for stimulating discussions. Work partially supported by PRIN INAF 2014 `Probing the AGN/galaxy co-evolution through ultra-deep and ultra-high-resolution radio surveys'.
\end{acknowledgements}

\clearpage

\clearpage

\clearpage

\clearpage

\clearpage

\clearpage
\begin{turnpage}
\begin{deluxetable}{lcccccccccccccccccccccccccccccccccc}
%\rotate
\tabletypesize{\scriptsize}\tablewidth{0pt}\tablecaption{SFR Function
Parameters}\tablehead{\colhead{Parameter} & &\multicolumn{7}{c}{Global (far-IR+UV)} & & &
\multicolumn{7}{c}{UV (dust-corrected)}\\
\\
\cline{1-1} \cline{3-9} \cline{12-18}\\
\colhead{} & &\colhead{$p_0$} & & \colhead{$p_1$} & & \colhead{$p_2$} & &
\colhead{$p_3$} & & & \colhead{$p_0$} & & \colhead{$p_1$} & & \colhead{$p_2$}
& & \colhead{$p_3$}} \startdata
$\log \mathcal{N}(z)$ & &$-2.47\pm 0.06$& &$-6.79\pm 1.19$& &$14.11\pm 3.63$& &$-9.58\pm 2.83$& & & $-1.95\pm 0.07$& &$-1.85\pm 1.57$& &$4.91\pm 4.04$& &$-5.73\pm 2.73$\\
$\log\dot M_{\star, c}(z)$ & & $1.24\pm 0.05$& &$5.37\pm 0.63$& &$-4.22\pm 1.87$& &$-0.71\pm 1.53$ & & &$0.01\pm 0.05$& &$2.5\pm 0.97$& &$1.38\pm 2.51$& &$-2.38\pm 1.69$\\
$\alpha(z)$& &$1.27\pm 0.01$& &$2.92\pm 0.23$& &$-6.47\pm 0.66$& &$4.48\pm 0.45$& & &$1.11\pm 0.02$ & & $2.88\pm 0.47$& &$-6.27\pm 1.21$& &$4.50\pm 0.78$\\
\enddata
\tablecomments{Quoted uncertainties are at $1-\sigma$ level. Fits hold in the range of SFR $\dot M_\star\sim 10^{-3}-10^4\, M_\odot$ yr$^{-1}$ and for redshifts $z\sim 0-10$.}
\end{deluxetable}
\end{turnpage}

\end{document}